# Context-Auditor: Context-sensitive Content Injection Mitigation

Faezeh Kalantari
Arizona State University
Tempe, AZ, USA
faezeh.kalantari@asu.edu

Mehrnoosh Zaeifi
Arizona State University
Tempe, AZ, USA
mzaeifi@asu.edu

Tiffany Bao
Arizona State University
Tempe, AZ, USA
tbao@asu.edu

Ruoyu Wang
Arizona State University
Tempe, AZ, USA
fishw@asu.edu

Yan Shoshitaishvili
Arizona State University
Tempe, AZ, USA
yans@asu.edu

Adam Doupé
Arizona State University
Tempe, AZ, USA
doupe@asu.edu

## ABSTRACT

Cross-site scripting (XSS) is the most common vulnerability class in web applications over the last decade. Much research attention has focused on building exploit mitigation defenses for this problem, but no technique provides adequate protection in the face of advanced attacks. One technique that bypasses XSS mitigations is the scriptless attack: a content injection technique that uses (among other options) CSS and HTML injection to infiltrate data. In studying this technique and others, we realized that the common property among the exploitation of all content injection vulnerabilities, including not just XSS and scriptless attacks, but also command injections and several others, is an unintended *context switch* in the victim program's parsing engine that is caused by untrusted user input.

In this paper, we propose Context-Auditor, a novel technique that leverages this insight to identify content injection vulnerabilities ranging from XSS to scriptless attacks and command injections. We implemented Context-Auditor as a general solution to content injection exploit detection problem in the form of a flexible, stand-alone detection module. We deployed instances of Context-Auditor as (1) a browser plugin, (2) a web proxy (3) a web server plugin, and (4) as a wrapper around potentially-injectable system endpoints. Because Context-Auditor targets the root cause of content injection exploitation (and, more specifically for the purpose of our prototype, XSS exploitation, scriptless exploitation, and command injection), our evaluation results demonstrate that Context-Auditor can identify and block content injection exploits that modern defenses cannot while maintaining low throughput overhead and avoiding false positives.

## 1 INTRODUCTION

Though rich, interactive, Web 2.0 applications are critical in enabling the modern web, they are also a critical attack vector on the Internet. Web application vulnerabilities have significantly contributed to the financial loss from cybersecurity issues over the past years. Among vulnerabilities in web applications, cross-site scripting (XSS) is the *single most common* type of vulnerability in the past two years according to Bugcrowd [20] and Hackerone reports [27]. In fact, since the release of the initial US-CERT Advisory in 2000 [57], XSS is among the most critical web application security threats every year, consistently appearing in "worst of" lists, such as the OWASP Top 10 [46] and MITRE top 25 [44].

Research into XSS prevention and mitigation has continued since XSS was first discovered. Existing approaches attempt to statically identify XSS vulnerabilities in server-side code [30, 34, 61], analyze the use of server-side sanitization functions [17, 40], filter out JavaScript code on the server side [23, 58], or attempt to detect the presence of vulnerabilities from the client's viewpoint [39, 43]. Yet, XSS vulnerabilities continue to manifest in web applications. A different class of solutions attempt to *mitigate XSS exploits* rather than detect the underlying vulnerabilities. These solutions include XSS filters in browsers (e.g., NoScript for Firefox [25]), Content Security Policy (CSP) [5], web application firewalls (e.g., ModSecurity [12]), and server-side HTML sanitizers [29]. These mitigations are widely adopted in practice (as we discuss in Section 2.2), however research has shown that it is possible to bypass these defenses.

One particularly interesting bypass of XSS mitigations, called scriptless attacks, generalizes the concept of XSS *beyond* the injection of JavaScript code [28]. Scriptless attacks inject data (such as CSS, HTML 5, SVG, and font files) to compromise the security of web applications, which, as a result, allows attackers to steal sensitive information, even in a restricted environment without JavaScript execution. Prior solutions that check if untrusted user input is used in sensitive output functions (e.g., `echo()` or `system()`) [31, 34, 38], or identify the mixing of code (HTML) and data (JavaScript) in the same channel [23], do not work for scriptless attacks, as the notion of code-data mixing is not fine-grained enough.

These exploit techniques inspired us to examine the root cause of XSS and scriptless attacks: Web applications embed into an HTML page untrusted user input as *pure data*, such as strings and text, with the intent that no part of the untrusted user input shall be interpreted by the parser as non-data. However, this important developer-intent is lost once the HTML code is generated and sent to the browser: client-side parsers must *re-discover* the meaning of every part of the page by parsing the HTML response. In carrying out XSS exploits, attackers exploit this loss of information and mislead client-side parsers to transition from HTML parsing to JavaScript or CSS parsing when parsing the user input, which *violates the developer's original intent*. For example, in an XSS exploit, an untrusted user input may cause a browser to transition from parsing plain text in HTML to parsing JavaScript code with a `<script>` tag. We term these context transitions in a parser *context switches*, where "context" refers to the functionality of the token being parsed. In fact, we argue that unintended context switches are the root cause of a series of vulnerabilities including (among others) XSS, scriptless attacks, command injection, SQL injection, XML injection, and template injection. We refer to them as *content injection vulnerabilities* throughout this paper.

Our insight is that, because these vulnerabilities share a common root cause, exploits against them can be mitigated using a common approach. In this paper, we introduce Context-Auditor, a novel, general technique that detects content injection exploits by identifying unintended context switches, caused by untrusted input, during parsing. This idea is inspired by Stock et al. [54], who used context switches (tokenization-based) in the JavaScript parser to detect DOM-based XSS exploits using a taint-tracking browser and by other string-based [33] and taint-based [52] injection prevention methods. Context-Auditor expands these ideas, generalizing these concepts to the broader category of content injection exploits.

We implemented multiple prototypes of Context-Auditor in various forms—including a shell wrapper, an nginx module, a web proxy, and a Chrome extension—to detect content injection exploits in shell commands, HTML, CSS, and JavaScript. We tested our prototypes on reputable testing suites and comprehensive real-world data sets, where Context-Auditor successfully detected and blocked all reflected XSS, scriptless, and command injection exploits in a number of web applications. Additionally, Context-Auditor demonstrated negligible false positive rate in a live crawl of the Alexa top-1000 websites.

Overall, this paper makes the following contributions:

- We reformulate the problem of defending against command injections, XSS, and scriptless attacks as a *content injection mitigation problem*, and we focus on the root cause of content injection exploitation, which is untrusted user input triggering a context switch in parsers.
- We build a custom parser model that supports HTML, CSS, JavaScript, and Bash scripts (and is extensible to other languages). Context-Auditor uses this model to identify context switches caused by untrusted user input, which indicates the exploitation of a content injection vulnerability.
- We demonstrate that Context-Auditor can mitigate exploits that state-of-the-art XSS mitigation techniques cannot, with low false positive rates and reasonable throughput overhead.

In the spirit of open science, we will open source Context-Auditor and publish the evaluation data and configurations to guarantee experiment reproducibility.

## 2 BACKGROUND

In this section, we provide an overview of content injection exploits that Context-Auditor aims to address along with an overview of existing mitigation techniques.

### 2.1 Content Injection Vulnerabilities

A web application that allows the usage of user data with unintended semantics in generating dynamic content is susceptible to content injection vulnerabilities. Therefore, an attacker can compromise the security of the web application by sending an exploit that leverages this vulnerability to perform developer unintended actions. For instance, if user input is used to construct a string that is then passed to a shell script command, an adversary can

```
1  <html>
2  <body>
3  <style>
4  body{
5    background-color: <?php echo($_GET["color"])?>
6  }
7  </style>
8  <form action="index.php">
9    background Color :</td><td>
10   <input type="name" name="color" />
11   <input type="submit" value="Change Color" />
12   <input type="hidden"
13     name="CSRFToken" value="SECRET">
14 </form>
15 <script>
16   document.write("Username is: ");
17   var str = "<?php echo($_GET["id"])?>" ;
18   document.write("<text>"+str+"</text>");
19 </script>
20 <h2> You were searching for:
21   '<?php echo($_GET['term']) ?>'
22 </h2>
23   Here is the result:
24   <?php
25   $command= 'cat userinfo.txt | grep '. $_GET['term'];
26   echo(exec($command)); ?>
27 </body>
28 </html>
```

**Listing 1: PHP code with three content injection vulnerabilities (Line 5, Line 17, and Line 21 represent CSS, JavaScript and HTML contexts respectively) in the server-side HTML response, one DOM-based XSS vulnerability on Line 18 and a content injection vulnerability in the form of command injection on Line 26.**

construct an exploit using shell script control characters, such as semicolon, to escape out of the current command and execute arbitrary commands. In another case, an attacker may alter the intended semantics of a web page or execute arbitrary JavaScript code if a web application uses user input to construct HTML response without proper sanitization.

**Web-based exploits.** We classify content injection exploits that manipulate the structure of an HTML response as *web-based exploits*, which are categorized into two groups: *scripting* exploits and *scriptless* exploits. Listing 1 is an example of vulnerable PHP code with three web-based content injection vulnerabilities (and DOM-based XSS) in different contexts (HTML, JavaScript, and CSS), and we will refer to it throughout the paper.

- **Scripting exploits.** Scripting exploits are a traditional and well-known attack vector that involve the injection of malicious JavaScript code into web pages. An example of scripting exploits is a cross-site scripting (XSS) exploit. Two of the vulnerabilities in the PHP code in Listing 1 can be exploited with scripting exploits. Both are reflected XSS vulnerabilities: One is in the HTML context on Line 21, and the other is in the JavaScript context at Line 17. The first vulnerability on Line 21 can be exploited by a classic HTML-context XSS exploit: `<script>alert('injection');</script>`. The second vulnerability on Line 17 can be exploited by a JavaScript-context XSS exploit: `Hi"; alert('injection');"There`.
- **Scriptless exploits.** Scriptless exploits are another type of content injection exploits where attackers embed into the DOM tree non-scripting elements (e.g., images or style sheets) that violate

```
1  "} a[href*='A'] {
2    background: url(attacker.com?A); } ...
3  a[href*='S'] {
4    background: url(attacker.com?S); } ...
5  a[href*='Z'] {
6    background: url(attacker.com?Z);}
7  a[href*='A'][href*='A']{
8    background: url(attacker.com?AA); } ...
9  a[href*='S'][href*='E']{
10   background: url(attacker.com?SE); } ...
11 a[href*='S'][href*='E'][href*='C'][href*='R']{
12   background: url(attacker.com?SECR); }...
```

**Listing 2: An example CSS context exploit used to exploit the scriptless content injection vulnerability shown in Listing 1.**

security policies in browsers, most notably, CORS policies [28]. Listing 1 has a vulnerability on Line 5 that can be exploited with a scriptless content injection exploit. Attackers can inject a malicious CSS context exploit, such as the example exploit in Listing 2, to leak the characters used in the secret Cross-Site Request Forgery (CSRF) token.

**Non-web-based exploits.** Content injection exploits happen not only in client-side web languages, but also in many other contexts, such as SQL queries and shell commands. For the `exec` function (Line 26 of Listing 1): the shell command includes a user-controlled string which enables an attacker to execute arbitrary shell commands: sending `Auditor; rm userinfo.txt` to the web application would cause it to remove the `userinfo.txt` file.

## 2.2 Mitigations of Content Injection Exploits

Generally, there are four types of mitigations for web-based content injection exploits (although thus far they have mostly focused on XSS exploits), and all are bypassable [38]:

**Browser-based XSS Filters.** Some browsers have built-in XSS detection and mitigation mechanisms to block malicious-looking HTML requests and responses. NoScript [25] (for Firefox) is a powerful browser XSS filter that is widely used. There is also a No-Script Chrome extension [60]; it could be used as a temporary XSS mitigation measure after the retirement of XSS-Auditor [18] from the Chrome browser. Detection capability of XSS filters is often limited to pattern matching solutions that are defined by regular expressions and they might become ineffective in the face of new exploits.

**HTML Sanitizers.** HTML sanitizers (e.g., DOMPurify [29]) are libraries used by web application developers to sanitize HTML text and filter (potentially malicious) content. The black-listing approach of sanitizers is usually very conservative, as characters are sanitized without a thorough understanding of their actual impact on the semantics of a web page. For instance, not all < characters inside HTML text are potentially harmful and need filtering. However, sending a string containing < character to DOMPurify, results in filtering of < and all of its following characters; although the original string might not eventually cause any security violation in the web application. Unsurprisingly, sanitization approaches cannot cover all possible content injection exploits (e.g., DOMPurify does not provide sanitization for JavaScript and CSS).

**Content Security Policy (CSP).** CSP is a white-listing mechanism that adds directives into HTTP headers or meta tags, which specify

```
1  GET /?id=Admin";alert(1);" HTTP/1.1
2  Host: vulnerable.com
```

**Listing 3: An HTTP request towards an nginx web server with ModSecurity enabled via OWASP core rule set (CRS).**

(among other things) the legitimate source of external resources that a web page can embed. In this way CSP, when used correctly, can mitigate many XSS exploits. While all modern browsers support CSP, it is not deployed properly by most web applications in the wild, and also previous work has shown it to be insufficient to prevent all XSS exploits [22].

**Web Application Firewall (WAF).** WAFs (e.g., ModSecurity [12]) attempt to detect malicious web requests and prevent them from reaching back-end web applications. WAFs need to be manually configured with a comprehensive set of rules or directives, comprehensibility of which correlates with WAFS detection capability. To demonstrate this, we configured the ModSecurity with the OWASP core rule set (CRS) [14], and it failed to detect a simple injection inside JavaScript code (Listing 3) against the web page of Listing 1.

## 3 OVERVIEW

The failure of state-of-the-art content injection mitigation techniques (Section 2.2) is due to prior approaches not addressing the root cause of context switching vulnerabilities. Parsers (e.g., HTML parser, JavaScript parser, shell parser, etc.) are the entities that eventually parse an exploit, yet they do not know if a context switch is triggered by attacker content or was developer intended. Therefore, in Context-Auditor, we suggest a fundamentally different approach: we model parsers using automata and detect any context switching caused by attacker-controlled input as a potential content injection exploit. Before discussing the details our context-switching-based detection approach, we first highlight the deficiencies of state-of-the-art techniques (motivating us to introduce a new detection approach) and the context switching concept. These prerequisites provide a high-level operational model of Context-Auditor.

## 3.1 Motivation

State-of-the-art mitigations attempt to identify characteristics of common exploits as malicious: NoScript and XSS-Auditor operate based on regular expression matching. Similarly, ModSecurity searches for known malicious-looking directives inside HTTP traffic, and DOMPurify identifies known potentially harmful characters or patterns. CSP only allows the inclusion of external files from the same domain as the origin. These black-listing (browser filters, ModSecurity, DOMPurify) or white-listing (CSP) approaches operate based on previously known patterns among content injection exploits, which makes them unprepared for unfamiliar exploits. Still, detecting the known patterns requires preparation of extensive regular expressions or a comprehensive set of directives, which is an error-prone and tedious task. More specifically, we argue that many mitigation techniques do not provide a comprehensive detection approach against content injection exploits because they fail to address content switches caused by untrusted input.

Bypass of ModSecurity (Listing 3) occurs as a consequence of context switching from the double-quoted string context in the JavaScript parser to after assignment, then to statement contexts. State-of-the-art techniques cannot detect context switches due to user-controlled data in web applications. Therefore, we recommend a fundamentally new content injection exploit detection approach in Context-Auditor that uses the context switching concept to detect content injections.

### 3.2 Context Switching

We define *context switching* as when a parsing engine changes parsing context (i.e., from one context to the next) based on the input. The parsing context is defined by the grammar rules of the specific language. For instance, consider the following grammar rule in ECMA262 (JavaScript) specification used to interpret double-quoted string literal. `StringLiteral` and `DoubleStringCharacters` are non-terminals in the language and `"` is a control character delimiating `DoubleStringCharacters` token from double-quote token:

```
StringLiteral :: "DoubleStringCharacters"
DoubleStringCharacter :: SourceCharacter but not one of " or \
```

A web application developer writes code that generates dynamic HTML and JavaScript content with implicit assumptions about the parsing of it as a `DoubleStringCharacter` token. However, if user-controlled input is used to generate a `DoubleStringCharacter` token without sanitization, it can violate the developer's intent by including a double-quote character. This violation constitutes a potential content injection exploit. We use this developer unintended context switching to detect a content injection exploit. We extend this concept to other languages that are parsing dynamic content of web applications.

Our focus on context switching as the root cause of content injection exploits allows us to provide a comprehensive and flexible detection technique. In Context-Auditor we model context switching for HTML, JavaScript, CSS, and Bash languages, and this model can be used as a stand-alone detection module with fine-grained content injection detection coverage.

### 3.3 Context-Auditor Overview

At a high-level, Context-Auditor operates as a stand-alone detection module as depicted in Figure 2. The high-level operational model of Context-Auditor is as follows:

**Input.** Context-Auditor analyzes the content generated by web applications to determine the impact of attacker-controlled input on parsing. The input to a web application takes the form of an HTTP request, and generated content includes HTTP responses or dispatched shell commands. As its input, Context-Auditor consumes the application's generated content and the location of untrusted (attacker-influenced) bytes in that content. If not given knowledge of untrusted bytes, we can infer the location of untrusted data in the generated content by identifying where content from the HTTP request is used verbatim inside the generated content[1].

**Parsing.** We designed automata to model context switches in the form of state transitions. These automata are made of states called *parsing states*: A parsing state includes syntactical and lexical information about the current character, which can determine the type of the token or the statement being parsed. These automata are also aware of the exact location in the nested structure of the language it is parsing. The parser is used to process the application output, and its design is detailed in Section 4.

**Detection.** The Context-Auditor automata parse the output content (HTTP response or shell command) using its parsing model from the *first character of the output* until the last character of untrusted (tainted) data. If a context switch occurs when parsing an untrusted data character in the application output, then Context-Auditor has detected an exploit. If no context switches occur as a result of untrusted data characters, then the untrusted input is benign.

**Mitigation.** When Context-Auditor detects an exploit, the user can choose to block the output from reaching its destination (i.e., does not deliver the HTTP response or does not allow the shell command to execute). This step can differ based on the user's deployment decision, and we discuss it in Section 5.

Figure 1 demonstrates the core idea of Context-Auditor, including simplified parsing automata of HTML, CSS, and JavaScript, along with the transitions between them. The bottom of Figure 1 has sample web application output (content input to Context-Auditor), and the arrows ⇑ indicate context switching of the parser on the index of the byte of the input (the transition on the edge of the parsing automata are labeled with the same index). Figure 1 also demonstrates two different scenarios. In the first scenario, the untrusted user input is `admin` (purple underline of the input), and the untrusted user input is identified based on the purple HTTP request. This input remains in the `Quoted Literal` state and does not cause any context switching, therefore it is benign. In the second scenario, the untrusted user input is `admin"; document.write(user);` (red overline of the input), and the untrusted user input is identified based on the red HTTP request. This input triggers a context switch at byte 56 (edge 56′) to `Stmt. End`, and because this context switch occurs while parsing untrusted user input Context-Auditor will detect this as malicious. As demonstrated, if untrusted user input does not cause a context switch (as is the case of the first scenario with `admin` as the untrusted user input in Figure 1). A similar situation exists when parsing shell commands: if untrusted user input is used as an argument, then it will not cause a context switch. In this way, Context-Auditor uses context switching to detect content injection exploits.

### 3.4 Vulnerability Model

Identifying where untrusted input occurs inside web application's dynamic content (e.g., HTML or shell commands), without knowledge of the server-side code, is a fundamentally difficult problem [21], and orthogonal to the detection of vulnerabilities by Context-Auditor (user-input offset is an input to Context-Auditor, as shown in Figure 2). Therefore, we outsource this orthogonal task to a module called `User-Input Detector`. Deployment of Context-Auditor inside an operational environment requires a `User-Input Detector` module in place, which in its simplest form performs string matching. To demonstrate the generality of Context-Auditor, we also deployed a heuristic from Buyukkayhan et al.'s work [21] as the User-Input Detector module in Section 6 (experiment E4): this heuristic identifies sensitive keywords of a payload and their counterpart inside an HTML response and then attempts to extend the matching offsets.

---
[1] We discuss the implications of this choice on Context-Auditor's efficacy in Sections 3.4 and 7.

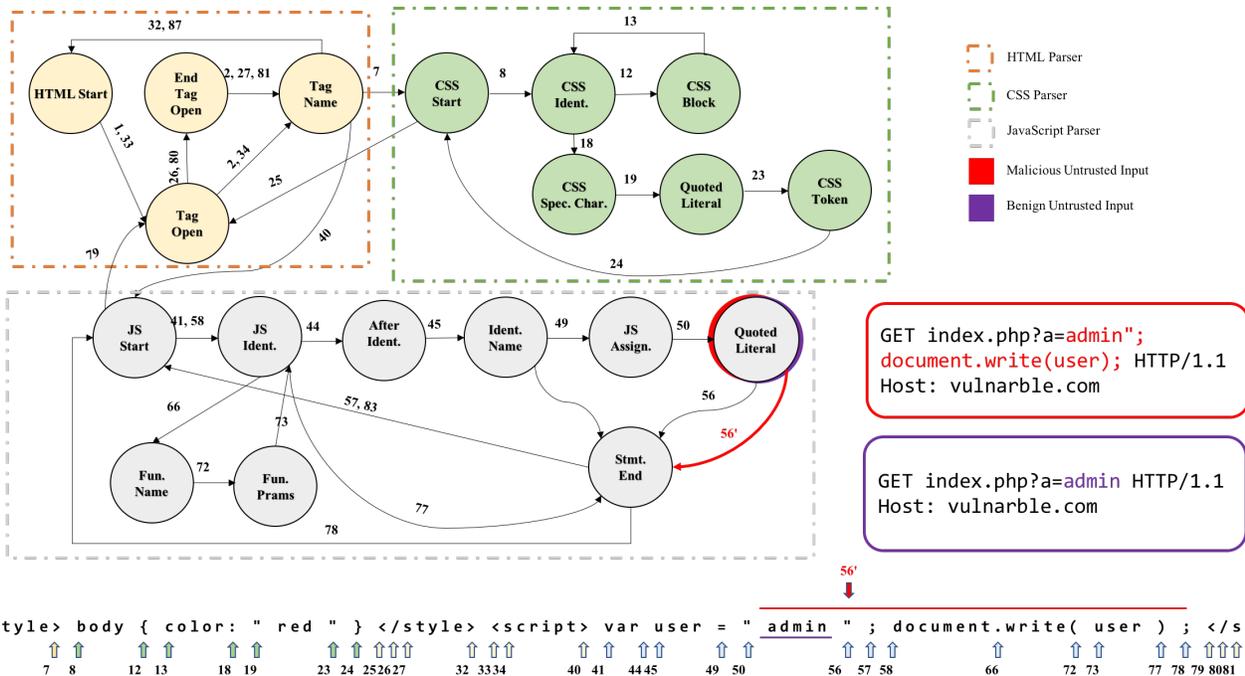

Figure 1: A simplified graph of context switching in a browser parser while parsing HTML input. If the untrusted input is `admin`, it does not trigger any context switch (parsing state is Quoted Literal for all characters of the input) and is therefore benign. However, if the untrusted input is `admin"; document.write(user);`, this triggers a context switch at edge $56'$ (from Quoted Literal state to Stmn. End state) and is therefore malicious.

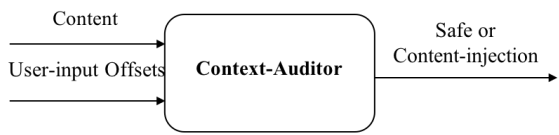

Figure 2: Context-Auditor is a detection black-box: given a specific content and byte offsets of untrusted input, it will return if the content is safe or is a content injection exploit.

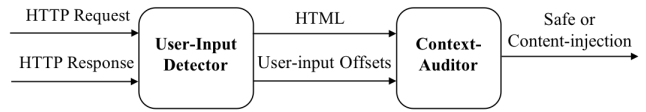

Figure 3: User-Input Detector identifies the location of untrusted input inside HTTP response and propagates inputs to Context-Auditor for further analysis.

As an example of deploying an instance of Context-Auditor as a detection module, we developed a prototype in which untrusted input (coming from the HTTP request) used verbatim in generated HTML content are identified and propagated to Context-Auditor for further analysis (Figure 3). In another deployment, User-Input Detector is a wrapper around /bin/sh and invokes Context-Auditor in cases that are susceptible to command injection exploits. These deployments have several implications on the types of vulnerabilities that can be protected against by our prototype. The prototype can prevent exploits against *non-stored* content injection exploits:

including reflected XSS, reflected scriptless exploits, and the vast majority of command injection exploits. If injected content is stored (for example, in a database) and later retrieved, our link of untrusted user input is lost. Without some additional tracking of this input (such as taint tracking and propagation), the prototype is not able to determine whether or not parsing context switches are caused by untrusted input.

The focus on reflected content injection allows Context-Auditor to operate without any knowledge about or modification to the underlying web application or web browser, and only concerns itself with the HTTP request and the content generated as a result of it. This allows Context-Auditor to be deployed in many configurations. For instance, similar to trusted-types [59] that uses DOMPurify to sanitize inputs to sensitive page's sinks (e.g. document.write and innerHTML_setter), we could integrate Context-Auditor in a browser to detect and limit unwanted context switching in sensitive sinks. Theoretically, it could also be deployed inside frameworks that detect stored XSS because these frameworks can track untrusted user input.

## 3.5 Threat Model

We designed our solution to protect against a sophisticated attacker. We assume that our attacker is aware of a content injection vulnerability in a target web application. The attacker can modify the content of any request parameter and send the request to the target web application to trigger the content injection vulnerability.

## 3.6 Comparison with Current Mitigations

Table 1 demonstrates the shortcomings of prior web-based mitigations against content injection exploits: XSS-Auditor only protects against conventional XSS vectors and does not support any exploit from HTML, CSS, or JavaScript contexts [54]. NoScript supports these three languages and prevents scriptless exploits via CSP rules; however its regular expressions might be ineffective in the face of new exploitation techniques or complex nested JavaScript exploits. DOMPurify is mainly focused on sanitization of HTML context and as mentioned in Section 2.2 it could be conservative in input sanitization. CSP is mostly concerned with inclusion of files or tags from external domains, and it cannot protect against injection of JavaScript and CSS code into existing JavaScript code. ModSecurity is limited to its directives, and it can be insufficient to prevent simple injections inside JavaScript code, scriptless exploits, and many HTML injections. However, Context-Auditor provides a fine-grained content injection exploit detection solution for scripting, scriptless, and command injection exploits (in HTML, JavaScript, and CSS contexts).

For command injection exploits, we referred to code and data separation solutions (e.g., SMask [31]) or taint-enhanced prevention policies [62]. They provide a content injection measure against trivial XSS and command injection, and both required some knowledge of server-side deployment. As discussed in Section 3.4, despite server-side (SMask and taint-enhanced policies), client-side (XSS-Auditor, NoScript, CSP, and DOMPurify), and WAF (ModSecurity) mitigation techniques that are tied to a specific location, Context-Auditor has a flexible deployment. is also advantageous for mobile browsers because they lag behind in implementation of similar security measures proposed for desktop browsers [42].

## 4 BUILDING THE MODEL

The parsing engine, in the form of an automaton, is the core of Context-Auditor. This automaton uses its state transitions to identify context switches, which is the key feature used to detect content injection exploits. We construct this automaton by manually analyzing HTML, JavaScript, CSS, and Bash languages.

### 4.1 Modeling Web Languages Parsing

To detect content injection exploits inside HTML content, we require an automaton to track parsing states of all characters in that content. If we design an automaton with precise parsing states encompassing syntactical and semantic information about the underlying token or character in each location, we would be able to detect a broad spectrum of content injection exploits that could be inserted into any location inside HTML content. Such automaton will consequently help us in the detection of exploits with various granularities: we can detect coarse-grained exploits that cause a language transition inside HTML content, and we can also detect fine-grained (and short) exploits that only insert additional functionality into existing code. Furthermore, unlike prior approaches, this technique requires no prior knowledge of exploits.

To design an automaton with the state transition requirements which supports three major web languages (HTML5, JavaScript, and CSS) and the Bash language, we studied specifications of these languages: specifically, we analyzed the lexical analysis stage from the HTML 5 specification [8] (which is referred to as the *tokenization stage*), the grammar rules from the ECMAScript specification ECMA-262 [2], a tokenization procedure introduced by W3C in CSS language specification [6], and the Bash language manual [4]. Considering the specifications, we realized that our candidate automaton must meet the following requirements:

**(1)** The next state of the automaton should be based on the current state and input character (or lexical token).

**(2)** The automaton should be able to track history to support nested structures, branches, arrays, objects, tags, etc.

**(3)** The automaton should support revisiting. According to HTML 5 parsing specification, the parser needs to re-consume a character under certain conditions [8]. Revisiting is also necessary for parsing CSS and JavaScript, as for each of them, certain characters in individual states might be the indication of either a new token or a new statement. In this case, the character must be re-analyzed in another state later which requires the revisiting property.

Based on these constraints, we design the automaton in Context-Auditor as a two-way finite pushdown automaton (2PDA): A pushdown automaton (PDA) is a finite-state automaton (FSA) with a stack, and a 2PDA complements PDA by supporting revisiting input characters that are already consumed. We designed our automaton in a similar fashion to the way that browsers parse an HTML document: the HTML parser is the primary parser shipped with any modern browser: It takes as input bytes representing an HTML document and starts parsing these bytes character by character. In the course of parsing, the HTML parser distinguishes among different tokens and special characters and even different languages in that document. For instance, while parsing a specific byte index, it determines whether the index is inside an opening tag, an attribute name/value or either inside a closing tag. The existence of special tags marking the beginning of style sheets or embedded scripts (e.g. `<script>` or `<style>` tags) is an indication of a new language for the HTML parser; it will invoke the corresponding parser (which is also included by the browser) to parse the embedded content between the opening and closing tags.

Our automaton starts parsing the HTML document from the first input character and then it moves forward character by character, while moving along the sequence, it has detailed information about the underlying language, token, and the specific language statement being parsed. We refer to such information as *context* and we use *parsing states* to reflect this concept inside an automaton. We manually constructed the HTML 5 parsing automaton following tokenization stage in the HTML 5 specification [8]. It tracks the HTML tokens and tags for each character in the HTML sequence via parsing states' information. It also supports transition into states that are related to CSS and JavaScript language contexts by reading a <script> or <style> tag. We also manually constructed parsing states related to CSS and JavaScript languages by referring to the related specification and grammar rules respectively (W3C tokenization specification for CSS [6] and ECMA-262 [2]). Figure 7 in Appendix shows a simplified representation of the 2PDA.

Table 1: Comparison of XSS mitigation techniques based on their detection capabilities and their applicability to mobile browsers. Missing detection capabilities are marked by numbers. 1: Depends on policy. 2: Pattern-matching-based. 3: Cannot detect complex nested JavaScript code. 4: It deploys CSP rules to prevent scriptless exploits. 5: Limited tags. 6: cannot detect content injected to CSS. In case of mobile support, we focused on whether the mitigation techniques introduced/considered any measures for content injection exploits that are sent from a mobile browser or not.

| Tool | Context-based Content Injection Exploit Mitigation | | | | Command Injection | Mobile Support | Location |
|---|---|---|---|---|---|---|---|
| | XSS | JS | HTML | CSS | | | |
| **XSS-Auditor** | ✓ | ✗ | ✗ | ✗ | ✗ | ✗ | client |
| **NoScript** | ✓ | Partial [2,3] | Partial [2,4] | Partial [2,4] | ✗ | ✓ | client |
| **DOMPurify** | ✓ | ✗ | ✓ | ✗ | ✗ | ✓ | client |
| **CSP** | Partial [1] | ✗ | Partial [5] | Partial [6] | ✗ | ✓ | client |
| **ModSecurity** | ✓ | Partial [2] | Partial [2] | ✗ | ✗ | ✗ | proxy |
| **SMask** | ✓ | ✗ | ✗ | ✗ | ✓ | ✗ | server |
| **Taint-enhanced policies** | ✓ | ✗ | ✗ | ✗ | ✓ | ✗ | server |
| **Context-Auditor** | ✓ | ✓ | ✓ | ✓ | ✓ | applicable | flexible |

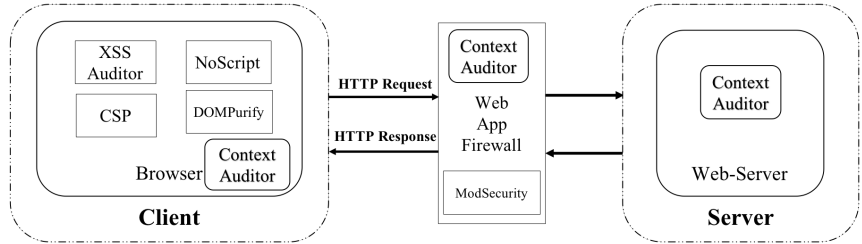

Figure 4: Possible deployment locations of all aforementioned XSS mitigations (XSS-Auditor, NoScript, CSP, ModSecurity, and WAF) and Context-Auditor in an HTTP client and server communication model.

### 4.2 Modeling Shell Command Parsing

Command injection exploits in web applications are the result of dynamic generation of shell commands using untrusted input. Similar to content injection detection for web languages, command injection exploits are also detectable via identifying context switching. To model shell command parsing, we focused on identifying different tokens in a shell command. We determine whether the current character is part of a command or operand, or if it is a special character which would change the type of a statement. The parser also considers the hierarchical structure of a command to keep track of quotes, back-ticks, parentheses, braces, etc. Figure 8 in Appendix shows a simplified 2PDA of the shell parsing automaton.

## 5 IMPLEMENTATION

As discussed in Section 3.2, our context switching-based content injection detection method provides us (among other benefits) with the advantage of flexible deployment. Despite dedication of other state-of-the-art content injection mitigation techniques to a specific location in the HTTP client–server communication model, as shown in Figure 4, Context-Auditor can be deployed anywhere along this model. In our experiments we deployed Context-Auditor in four different operational models:

**Context-Auditor Shell Wrapper:** Figure 5 shows how Context-Auditor works on server side as a module to detect command injection exploits. This module is a wrapper around /bin/sh that implements the functionality of User-Input Detector: It generates byte offsets marking reflections of HTTP request parameters in a series of commands sent to /bin/sh and invokes Context-Auditor's shell parser module with the input commands and byte offsets.

**Context-Auditor nginx Plugin:** Figure 6 illustrates how Context-Auditor works in an environment with nginx as the HTTP frontend, Apache httpd (or other web applications hosted by Apache httpd) as the service back-end, and Context-Auditor as a standalone detection module. The nginx plugin acts as an HTTP proxy and detects untrusted input by identifying any input data from the HTTP request in the HTML response. Then, nginx forwards the intercepted data to Context-Auditor, and, after receiving the detection results from Context-Auditor, the nginx plugin filters and blocks all responses that are content injection exploits.

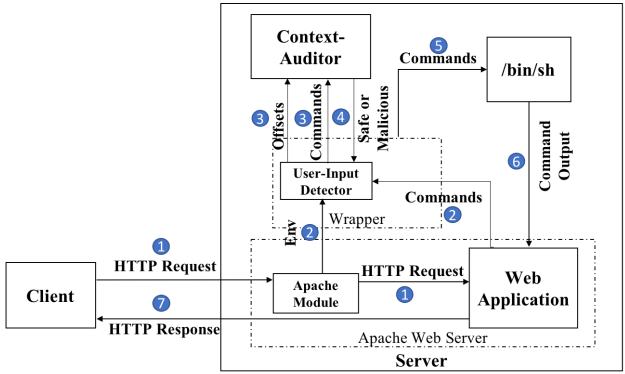

Figure 5: The communication diagram of a Context-Auditor-equipped HTTP client-server model to prevent command injection exploits.

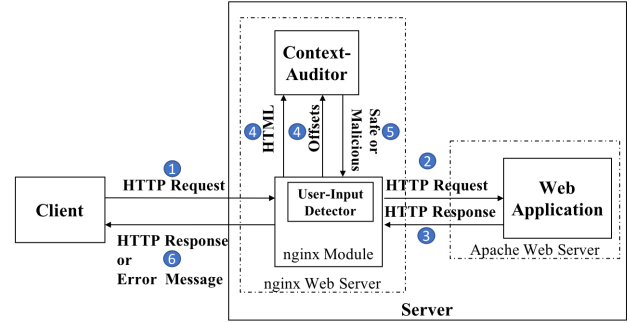

Figure 6: The communication diagram of a Context-Auditor-equipped HTTP client-server model to prevent injections inside a HTTP response.

**Context-Auditor Web Proxy:** We integrated Context-Auditor with *mitmproxy* [9]. This instance identifies byte offsets of user-controlled data in HTTP responses and invokes Context-Auditor for parsing analysis of the response. If a potential content injection exploit is detected, then Context-Auditor blocks the response.

**Context-Auditor Chrome Extension:** We modified an existing Chrome extension called *Tamper* [10], which allows us to intercept Chrome's HTTP requests and responses. From the request and response, we identify untrusted input and pass this to Context-Auditor to detect content injection exploits. Due to technical limitations of the Tamper extension, we cannot block responses and log them instead.

## 6 EVALUATION

In evaluating Context-Auditor, we sought to answer the following research questions:

**Q1. Effectiveness.** How many content injection exploits in HTML, CSS, JavaScript, and shell scripts can Context-Auditor detect, and how many exploits does Context-Auditor miss?

**Q2. Practicality.** Does Context-Auditor exhibit a low false positive rate so that it can be deployed in real-world settings without raising excessive false alarms?

**Q3. Efficiency.** Does Context-Auditor exhibit a low runtime overhead in all evaluated scenarios to justify the deployment of it in real-world settings?

### 6.1 Experiment Desgin

We designed seven experiments under different scenarios with different data sets. Two experiments (E1 & E2) focus on the true positive rates of Context-Auditor in detecting XSS and scriptless exploits. To stress the generality of our technique, we also designed an experiment (E3) demonstrating the effectiveness of Context-Auditor in detection of command injection exploits. E4 showcases the prevalence of the context switching phenomenon in a public data set of XSS exploits. Another three experiments (E5, E6, and E7) focus on showing the practicality (false positive rates) and efficiency (runtime overhead) of Context-Auditor in real-world settings. Table 2 shows an overview of these experiments.

### 6.2 Data Sets

We used both well-known public data sets and hand-crafted data sets in our experiments. Public data sets include the PortSwigger cross-site scripting cheat-sheet (E1) [56], Buyukkayhan et al.'s [21] data set of reflected server-side XSS exploits (E4), and the OWASP XSS cheat-sheet (E7) [3]. These are all ground-truth content injection exploits that should be detected and blocked by a perfect XSS defense. We ran Context-Auditor on these data sets to evaluate its effectiveness. Additionally, as discussed in Section 2.2 and Section 3.6, we provide hand-crafted XSS exploits that bypass several state-of-the-art XSS defenses but can be detected by Context-Auditor, which we will not reiterate in this section. Because of the lack of publicly available data sets, we also built our own data sets of command injection exploits (E3) and web pages that are free of content injection exploits based on known CVEs and a crawl of top Alexa websites (E5), respectively.

### 6.3 E1: Detecting Web-based Content Injection Exploits

To evaluate the effectiveness of Context-Auditor against real-world web-based content injection exploits, we ran our prototype against public data sets of XSS and scriptless exploits from PortSwigger XSS cheat-sheet [56]. Portswigger website provides URLs via link to vulnerable web applications for every exploit; we used both vulnerable web applications and exploits from the website in the experiment. We selected reflected XSS (including event handlers, consuming tags, file upload, restricted characters, frameworks, protocols, special tags, other useful attributes, encoding, obfuscation, and WAF bypass categories) and scriptless exploits specified to work on the Google Chrome browser. The exploits also cover both HTML and JavaScript contexts. Then, we use Chrome to access each URL, proxy the requests through the Context-Auditor Web Proxy (as described in Section 5), and report the number of detected exploits. Reflected exploits constitute the majority of exploits in the Portswigger dataset; therefore, we configured the experimentation environment via a User-Input Detection module that identifies request parameters used verbatim inside their consequent HTML response. Context-Auditor demonstrated 100% detection rate: It successfully detected all 242 XSS exploits and all 25 scriptless exploits.

Table 2: Details of each experiment in the evaluation. "Standalone" means Context-Auditor is used as a standalone Python library. The other four types of deployment are described in Section 5.

| ID | Description | Category | Deployment | Data set | Related Question |
|---|---|---|---|---|---|
| E1 | XSS and scriptless exploits detection | web | web proxy | PortSwigger XSS cheat-sheet [56] | Q1 |
| E2 | XSS detection, comparing against w3af | web | nginx plugin | Firing Range [7] | Q1 |
| E3 | Command injection exploits detection | shell | shell wrapper | web applications with known command injection CVEs | Q1 |
| E4 | Context switching probability in a public data set | web | standalone | data set from [21] including exploits from xxsed [24] and obb [11] | Q1 |
| E5 | False alarms and runtime overhead | web | web proxy | Alexa Top 1,000 websites | Q2 and Q3 |
| E6 | False alarms | web | web proxy | WordPress and human-generated traffic | Q2 |
| E7 | Runtime overhead | web | nginx plugin<br>web proxy<br>Chrome extension | hand-crafted data set | Q3 |

## 6.4 E2: Detecting XSS Exploits Generated by W3af

In this experiment, we deployed Context-Auditor as an nginx plugin (Figure 6) and tested if it can prevent XSS exploits coming from attackers. We first hosted under nginx 32 web applications provided by Firing Range[2] [7] that have reflected XSS vulnerabilities. The content injection vulnerabilities in Firing Range may occur in all three contexts, which are HTML, CSS, and JavaScript. We then ran the open-source w3af web vulnerability scanner [13] against these web applications, and w3af successfully detected XSS vulnerabilities and generated XSS exploits on the 32 web pages. Then, we reset the deployed web applications, enabled the Context-Auditor nginx Plugin (as described in Section 5), and re-ran w3af against these applications. Context-Auditor correctly detected and blocked all requests with exploit payload generated by w3af, without blocking any benign requests. As a result, w3af did not report any XSS vulnerabilities. This shows that Context-Auditor can detect and block realistic XSS exploits.

## 6.5 E3: Detecting Command Injection Exploits

Due to the lack of existing data set of web applications with command injection vulnerabilities, we manually scanned all CVEs to build a data set that comprises PHP web applications with known command injection vulnerabilities. As dictated by the threat model (Section 3), Context-Auditor only supports the situation where the exploit payload is sent as a URL parameter and used verbatim in a shell command. From these, we sampled three vulnerable web applications. We additionally checked a vulnerability report published by RIPS [1] vulnerability scanner's website and added another vulnerable application to our list. With these four web applications, we verified the effectiveness of the Context-Auditor Shell Wrapper (as described in Section 5) to detect command injection exploits when it is deployed in a scenario similar to Figure 5. Context-Auditor successfully detected content injection exploits for all of these applications. Table 3 shows the list of these vulnerable web applications and the detection capability of Context-Auditor.

---

[2]Firing Range is a web application test suite that contains a wide range of intentional vulnerabilities. We used version 0.48 in this experiment.

Table 3: Context-Auditor's command injection detection capability. 1: Denotes multi-token commands.

| CVE | Application | Version | Detection |
|---|---|---|---|
| **CVE-2015-5958** | PHP File Manager | 0.9.8 | Partial [1] |
| **CVE-2010-4278** | Pandora FMS | 3.1 | ✓ |
| **CVE-2008-6669** | nweb2fax | 0.2.7 | ✓ |
| **0-day by RIPS [1]** | Oscommerce | 2.3.4 | ✓ |

The command injection vulnerability in the PHP File Manager resides inside a backdoor that allows an attacker to execute arbitrary OS commands, which is an intended malice. Because malicious intention in a web application is outside the threat model of Context-Auditor, whenever an attacker executes a single token command (such as *whomai* or *ls*), Context-Auditor does not detect it. However, Context-Auditor *can* block any command with more than one token sent to the web application.

## 6.6 E4: Measuring the Likelihood of Context Switching in Exploits of a Public Data Set

Context switching concept is the core idea of Context-Auditor, to realize the importance of it, we analyzed its prevalence on a public data set of XSS exploits. We used a data set created by Buyukkayhan et al. [21]: They performed a longitudinal study on reflected server-side XSS exploits from XSSED [24] and OPENBUGBOUNTY [11] data sets and consolidated those in their data set. This data set has an attack table including data from actual exploits; such data involves payloads and exploited HTML responses. However, each payload could have several reflections in its HTML response; all of those might not lead to actual exploits. They implemented a greedy heuristic to identify these candidate reflections (user-input detector module) and then used a series of methods to extract a single working exploit from each response. Their heuristics was a proper user-input detector module, therefore we reached out to the authors and gained access to the heuristic's code. Afterwards, we ran an analysis on Buyukkayhan et al.'s [21] data set: For each payload and HTML response in the data set, we ran them through the heuristic and received a list of candidate exploit offsets. Context switching idea claims that at least one of these candidates triggers a context switching; therefore, we aimed to verify this claim in this experiment. In this regard, we analyzed the parsing of 170,667 entries of the data set. Context-Auditor successfully flagged 148,778 (87.17%) of those payloads that trigger a state transition in an HTML response; but, it did not report any state transition for 21,889 of the entries (12.82%) in the data set. We manually investigated some of these payloads to understand their nature. The inability of Context-Auditor in handling such exploits is for the following reasons:

(1) Context-Auditor is a research prototype to demonstrate the practicality of the context switching concept in detecting content injection exploits. Since we have manually designed and implemented our parsing automata, it is prone to parsing errors and incompleteness. We devoted significant engineering effort to improve its coverage and detection capability, however, there is still room for the implementation of methods to cover more corner cases and inputs via syntactical errors. (2) Some payloads do not trigger a context switching in any of the offsets. Such cases might occur due to the reflection of an exploit inside an HTML attribute value such as `onclick`, or it might lead to a second-order XSS exploitation. Our current implementation of Context-Auditor could detect exploits that force the HTML parser to transfer (from `attribute name` state) into `attribute value` state; but it does not support interpretation of JavaScript context inside certain `attribute value` states (e.g. `onclick`'s value). However, this is not an inherent limitation of our approach, and it is a matter of engineering effort. In the case of second-order XSS exploits, if the exploit reflects inside the assignment part of a JavaScript `Quoted Literal` for instance, it does not trigger a context switching at the time of parsing. But, it might lead to exploitation at runtime, which Context-Auditor fails to detect. Such exploitations are not in the scope of our experimentation setup, we could theoretically detect a few of those by deploying Context-Auditor at sensitive JavaScript sinks similar to Trusted Types [59] and defining injection detection policies (similar to [47]).

### 6.7 E5: Performance Overhead and False Positive on Top Alexa Websites

To evaluate the performance overhead and false positive rate of Context-Auditor in real-world settings, we performed a crawl of the top 1,000 Alexa websites [16]. For each URL we crawled, we followed two random links (links with at least one parameter) and measured the average loading time (over two runs) for the random links. Since Context-Auditor is not invoked for the main URLs (due to lack of URL parameters), we only considered the loading time overhead of random links and excluded the overhead of main links. Additionally, we monitored any potential false alarm triggered by Context-Auditor in this experiment. Similar to E1, we use Chrome to access each URL and proxy the requests through mitmproxy (as described in Section 5). Since the experiment involved benign traffic, we used a User-Input Detector module that recognizes request parameters used verbatim inside their consequent HTML response, and Context-Auditor is invoked if the length of those reflections is at least three characters. In case there are multiple reflections of the URL parameters inside the HTML response, we ran Context-Auditor for up to five reflections of each URL parameter value.

Since mitmproxy imposes significant overhead itself, we first crawled the random links in an experimentation setup with mitmproxy (Context-Auditor disabled) deployed, then we revisited the links in an environment with both mitmproxy and Context-Auditor enabled. We filter out any request timed out (with a timeout of 20ms per request). Also, we did not involve links with negative loading time overhead in our loading time measurements. Our measurements demonstrate an average of 4.7 seconds in loading time overhead while visiting random links that involve an average of three URL parameters. Impressively, Context-Auditor only flagged one of the visited random links as malicious. The loading time is caused to the following reasons: (1) In this experiment Context-Auditor analyzes up to five reflections for each request parameter value. For each, it parses the HTML response from the first character, which causes excessive delays. (2) The parsing automata cannot correctly handle some syntax errors or parsing of JavaScript statements not being delimitated by semicolons. We implemented a monitoring algorithm (timing-based) that detects such cases; it then adjusts the JavaScript code (by semicolon insertion) or forces the parser into particular states, all of which impose delays.

### 6.8 E6: Measuring False Positives on a Blog Site

Context-Auditor may theoretically produce false positives when user input causes context switches in the dynamic content of a web application, and this action is part of the expected functionality of the application. An example is a blogging website where users may embed JavaScript code into their blog posts. To understand the severeness of this scenario in a real-world setting, we performed a study on the WordPress platform, a widely used blogging web application, with a set of 10 human testers. We first deployed a WordPress instance on a server and put it behind a Context-Auditor web proxy. Then, given a unique user account and the URL to our WordPress instance, we allotted each tester 10 minutes and asked them to freely use WordPress and report any unresponsiveness of the application to us. Testers could create, add, delete, or modify any blog posts and input any data, including JavaScript code or CSS.

Throughout the experiment, testers did not report any case of unresponsiveness. Finally, a total of 1,680 HTTP request and response pairs were collected and analyzed by Context-Auditor. Context-Auditor did not report any content injection exploit, which means it did not raise any false alarms. We believe this is because normal WordPress users rarely need to insert custom JavaScript or CSS code into their blog posts despite the support in WordPress. Advanced WordPress users who need to insert JavaScript or CSS code into their blog posts may choose to ignore Context-Auditor warnings temporarily. Therefore, we argue that Context-Auditor can be applied in real world without causing excessive false alarms. In the unlikely case that a tester might have attempted to exploit WordPress, we further verified that these HTTP requests and responses do not contain any content injection exploits. So Context-Auditor also did not cause any false negatives in this experiment.

## 6.9 E7: Measuring Runtime Overhead

To evaluate the performance overhead of Context-Auditor, we used the Selenium web driver [15] to request webpages. We extracted a list of content injection exploits from the OWASP XSS cheat-sheet [3] categorizing them based on context. Then, we chose a proper vulnerable web page from the Firing Range application [7], for each exploit category and crafted corresponding links as the input to Selenium driver. We measured the time of fetching a webpage (1) without Context-Auditor, (2) with Context-Auditor as an nginx plugin, (3) with Context-Auditor as a proxy, and (4) with Context-Auditor as a browser extension. For the nginx plugin and the proxy, Context-Auditor returns a 404 response code when it detects a malicious request. This means that, depending on the deployment scenario, the latency to load the HTTP request may in fact, be faster for malicious requests. In case of the browser extension, the extension does not filter malicious responses, instead it only logs requests/responses including a content injection exploit.

Table 4 has the performance measurements of this evaluation. In the best case for benign requests, the Context-Auditor nginx module, added 4ms (27%) of overhead. The worst case, the Context-Auditor extension, added 19ms (127%) of overhead. The best case of malicious requests, the Context-Auditor nginx module, added 1ms (6%) of overhead, while the extension added 13ms (81%) of overhead in the worst case. As all implementations of Context-Auditor are proof-of-concept research prototypes (Context-Auditor is implemented in Python), we did not focus on optimizing latency.

## 7 LIMITATIONS

As we showed in our evaluation, Context-Auditor is a new approach to defend against content injection exploits. However, it does have a number of limitations, which we discuss in this section.
**False positives.** There was one false positive in our experiments, however some circumstances could cause more: For instance, if a blogging application (that allows blogging in raw HTML) reflects the user's newly created post (which is sent in as an HTTP parameter) in its response, Context-Auditor might report an exploit despite this behavior being legitimate and intended by the website's developers. This is not as common: blogging platforms typically send an HTTP redirect to the HTTP post request rather than reflecting the update itself, in which case there would be no false positive as the request will not be contained in the response. However, even if the platform does not use a redirect (and actually reflects the blog post), and Context-Auditor blocks the response, it will only block the response that includes the reflection: in this case, the submitted request will still be evaluated by the web application and the behavior would be correct (i.e., the blog post would appear).
**Second-order content injections.** We define second-order exploits as content injection exploits that do not trigger any parsing state transitions, yet still execute JavaScript. DOM-based XSS vulnerabilities are an example of second-order exploits: The content of an untrusted JavaScript string is interpreted—*during JavaScript execution*—as JavaScript code. These types of vulnerabilities are also called client-side XSS vulnerabilities, because the root cause of the vulnerability exists in the JavaScript code of the web page. In other words, there is no way to change the server-side code to fix the vulnerability (e.g., when untrusted input is used as an argument of eval function). Listing 1 contains a client-side XSS vulnerability on Line 18. The malicious input for the id parameter of the (JavaScript context) string literal <script>alert(1)</script>, will *not* cause a context switch. However, at runtime the browser's JavaScript execution engine will send the str variable (which is now untrusted data) to the document.write function, where it will interpret this string as HTML (thus causing a second-order content injection, where this parsing of the untrusted data by document.write will cause a context switch in the HTML parser). Unfortunately, Context-Auditor in its current deployment cannot detect such exploits. Mitigating second-order (and, perhaps *n*-order) content injection exploits would require modifying the browser to track the untrusted data (e.g., via taint propagation), similar to the approach proposed by Stock et al. [54]. However, modifying the browser is outside the scope of this work.
**Transformations of user input.** Server-side code or client-side scripts may sometimes perform transformations on user-controlled input, yet our current deployment of Context-Auditor supports cases where input strings are reflected verbatim in response data. However, using a different User-Input Detector module such as the heuristic from Buyukkayhan et al. [21]'s work that allows a few character mismatches between a payload and a candidate offset in the HTML response, Context-Auditor could detect more cases of context-switching. The tracking of input through transformations is a generally complex problem. There *are* approaches, such as dynamic taint tracking, that could be used to track such transformations. However, as mentioned in Section 3.4, the task of user input tracking is orthogonal to the core concept of Context-Auditor. Any increase in the ability of untrusted input tracking, regardless of the techniques used, would increase the space of content injection vulnerability instances that Context-Auditor can protect.
**Stored content injection.** As discussed in Section 3.4, Context-Auditor only supports non-stored content injections. This is due to Context-Auditor's inability to track untrusted input through data stores used by an application. This is an analogous issue to the transformation of user input: whereas that problem deals with the tracking of input through transformations, this must tackle the tracking of input through storage. One potential direction to address this limitation is through the use of proxies between the web application and its data stores: a very similar approach to SQL parsing analysis by libinjection [26], except that having a parsing automaton for query languages (an extension of Context-Auditor) we do not need to instrument web applications' source code, and we can delegate the injection detection task to a proxy. This proxy must implement a way of relating stored content and its corresponding HTTP request content (similar to how the /bin/sh wrapper is handled for shell injection) so that it could inform Context-Auditor of content containing untrusted input. This represents a significant engineering burden, but would expand Context-Auditor to yet additional content injection vulnerability classes.

Table 4: False negative and true positive rate of Context-Auditor with benign and malicious sets, along with performance analysis in terms of Latency (loading time in milliseconds) for the two lists, in four cases: without Context-Auditor or with any of three instantiations of Context-Auditor (nginx module, web proxy and Chrome extension).

| Case | Loading time without CA | CA loading time for nginx | CA loading time for proxy | CA loading time for extension |
| --- | --- | --- | --- | --- |
| **Benign Requests** | 16ms | 20ms | 17ms | 35ms |
| **Malicious Requests** | 15ms | 16ms | 18ms | 28ms |

**Syntax errors.** Context-Auditor is a research prototype manually built based on languages' grammars and specifications. We faced issues while parsing malformed HTML and JavaScript codes, which is the cause of some parsing failures in E4 (Section 6). We investigated numerous parsing errors and failures, resolving many of those. For instance, malformed JavaScript code or developers' negligence to deliminate statements via a semicolon could create syntax errors or indefinite parsing loops. Therefore, we implemented monitoring code to identify such cases and force the parser to Syntax_Error or Automatic_Semicolon_Insertion states when necessary. Another monitoring code modifies the JavaScript parser's input according to *automatic semicolon insertion* rules of ECMA-262 [2] specification when parser enters Automatic_Semicolon_Insertion state.

**Browser quirks.** Attackers may utilize browser-specific parsing quirks when carrying out exploits. These quirks are notoriously difficult to model, and Context-Auditor's parsing model certainly does not handle all of them. However, Context-Auditor will report an exploit as long as the attacker's input triggers either a parsing context transition or a parsing error. In the presence of an attack utilizing a browser parsing quirk, the former case would represent Context-Auditor correctly handling the quirk, and the latter would represent the common case of Context-Auditor *not* handling the quirk. In either case, the attack will be detected. Of course, some browser quirks that do not trigger parsing errors almost certainly exist. Context-Auditor would not be able to detect the exploits, and this is a limitation of our approach[3].

**XS-Leak.** Deployment of Context-Auditor in blocking mode might raise the XS-leak issue: In some instances, an attacker might infer the state of a victim at a target site by sending requests via specially-crafted inputs [55]. Here, Context-Auditor blocks the response since it can not distinguish between offsets that originate from an attacker and the ones that are already part of the HTML content. We can limit the XS-leak issue via the integration of taint-tracking approaches and only flagging context switching cases that are triggered by untrusted user input.

## 8 RELATED WORK

There are many research projects on detecting XSS exploits, such as sanitization [17, 40], policy enforcement [32, 62], code-data separation [23], taint tracking [58], moving target defense [45], and other server-side [19, 34, 41] and client-side [39, 43] mitigations. However, none of them tackles the root cause of content injection vulnerabilities. As our paper is based on context-sensitive parsing, we will only discuss papers with relevant approaches in the rest of this section. Stock et al.'s approach [54] in detection of DOM-based XSS vulnerabilities is the closest to our idea. They focus on the tokenization process with an observation that benign user input should only be tokenized into literal tokens, and any non-literal token coming from user input indicates an exploit. Context-Auditor generalizes their insight into the root cause of all content injection vulnerabilities and build a generic defense against such exploits. Prokhorenko et al. propose the context-centric injection detection model to identify exploits including XSS and SQL injection [48]. Compared to our automaton, their view of context is more coarse-grained. Their model is also tied to server-side PHP code.

---
[3]In fact, this is a fundamental limitation that man-in-the-middle components cannot always correctly infer all behaviors of client or server side components [49].

ScriptGard focuses on correct placement of sanitizer functions in server-side code considering the context that a sanitizer function is used in [50]. XSS-GUARD [19] uses parsing analysis to determine authorized scripts in an HTTP response that are intended by its developers. It introduces the concept of "Shadow Web Pages" and forces the application to follow the same execution path for both untrusted and benign input (by building parse trees and equivalency checks). Although their idea seems similar to Context-Auditor, the parsing state in our work exhibits a broader view by looking deeper into the syntactical structure of all supported languages. Moreover, our solution supports a wide range of content injection exploits instead of merely XSS exploits. Mitigation of content injection vulnerabilities is also beneficial to limit compromises caused by recent JavaScript-based vulnerabilities: Such as vulnerabilities in postMessage event handlers [53], client-side CSRF [36], prototype pollution [35], XS-leak [37, 55], and browse-based side-channels [51].

## 9 CONCLUSION

For too long, the research community has focused on mitigating XSS exploitation by blocking or detecting JavaScript execution. We believe that a shift in thinking is necessary: by broadening our scope to content injection, in terms of both vulnerabilities and exploits, we can finally address the root cause of content injection exploits: untrusted input that causes context switch in a parser. By modeling the parsing process, we can detect these exploits. This paper describes Context-Auditor, a generalized detection mechanism for content injection exploits. While Context-Auditor is capable of detecting a wide range of injection vectors, our prototype supports injections in HTML, CSS, JavaScript, and shell commands. Our evaluation showed that Context-Auditor is effective, performant, and unintrusive. Context-Auditor represents the first step to mitigating first-order webpage-based content injection exploits that can be applied on the server side, in a proxy, or on the client side. While Context-Auditor has pushed forward the state-of-the-art in content injection mitigation, more research remains to extend this idea to second-order content injection vulnerabilities—or even beyond the web.

## 10 APPENDIX

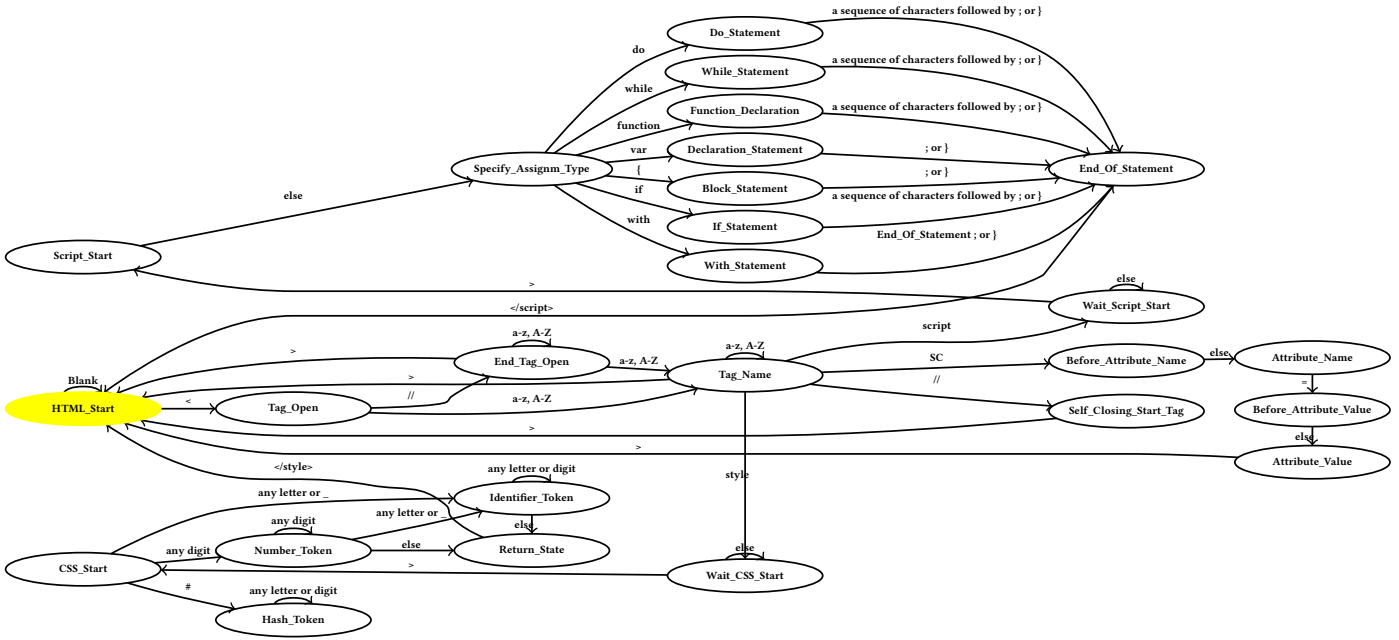

Figure 7: A simplified 2PDA that represents state transitions between HTML, CSS, and JavaScript. `HTML_Start` is the starting state. `Script_Start` and `CSS_Start` represent the states where first characters of JavaScript and CSS code are consumed by the automaton, respectively. The automaton transitions back to `HTML_Start` from JavaScript or CSS states whenever it consumes a sequence of characters that represent an end tag in the corresponding language.

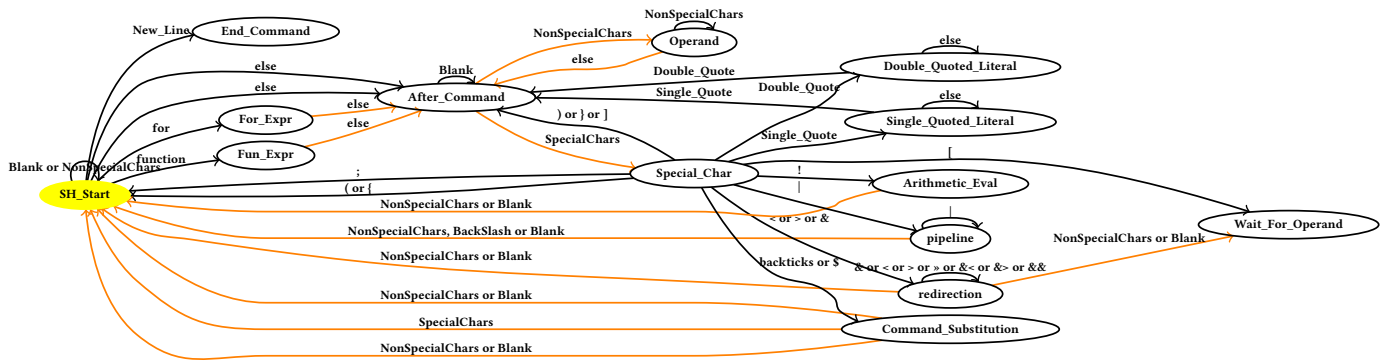

Figure 8: A simplified 2PDA that parses shell scripts. `SH_Start` is the starting state. Orange edges demonstrate the revisiting functionality.